# Energy considerations for a superlens based on metal/dielectric multilayers

Mark J. Bloemer,[1,*] Giuseppe D'Aguanno,[1] Michael Scalora,[1] Nadia Mattiucci,[1] and Domenico de Ceglia[1,2]

[1] *Department of the Army, Charles M. Bowden Research Facility, Aviation and Missile Research, Development, and Engineering Center, Redstone Arsenal, AL 35898*

[2] *Dipartimento di Elettrotecnica ed Elettronica, Politecnico di Bari, Via Orabona 4, 70125 Bari, Italy*

[*]*Corresponding author: mark.bloemer@us.army.mil*

**Abstract:** We investigate the resolution and absorption losses of a Ag/GaP multilayer superlens. For a fixed source to image distance the resolution is independent of the position of the lens but the losses depend strongly on the lens placement. The absorption losses associated with the evanescent waves can be significantly larger than losses associated with the propagating waves especially when the superlens is close to the source. The interpretation of transmittance values greater than unity for evanescent waves is clarified with respect to the associated absorption losses.

**1. Introduction**

Several years ago Pendry[1] showed that a simple metal film could be used as a lens to resolve an object beyond the diffraction limit. This phenomena works for TM polarized light and is based on two features: negative refraction of the Poynting vector which focuses the propagating waves and excitation of surface plasmons to extend the range of the evanescent components of the source.

The transmittance through a metal film is quite low and decreases exponentially with the thickness of the metal film. The low transmittance of metals limited their operating range to wavelengths in the vicinity of the plasma frequency (where metals are semitransparent) and to metal films of thickness ~50 nm or less. To overcome the low transmittance limitations of single metal films, a multilayer metal/dielectric stack was proposed [2]. These multilayer lenses do not focus the propagating waves but instead channel the light in the forward direction much like an optical waveguide. The transmittance can be improved with multilayers but the original proposal was based on very thin layers (5 nm thick) which are difficult to fabricate. Later designs incorporated thicker metal and dielectric layers [3-5] which are easier to fabricate. Although fabrication restrictions were eased by the thicker individual layers, low values of transmittance for propagating and/or evanescent waves remained an issue.

Recently a variation on the metal/dielectric superlens structure was proposed [6] which provides broadband transmittance for both the evanescent and propagating waves. The basic design in Ref. 6 is a stack of strongly coupled Fabry-Perot cavities with an antireflection coating on the entrance and exit faces. These one-dimensional metal/dielectric photonic band gap (1D MDPBG) structures are referred to as "transparent metals"[7-9] due to their highly transmissive window (>50%) across the visible spectrum and low electrical sheet resistance (~0.1 ohm/sq).

While the absorption losses for propagating waves in a 1D MDPBG have been described and follow the well-known relation of Absorption+Transmission+Reflection=1, the absorption losses for the evanescent waves have not been studied. It is known that the transmittance for evanescent waves can be >100% but the relationship of this impressive transmittance to absorption losses needs to be clarified. In the following we theoretically investigate the losses for these evanescent waves and examine what, if any, affect these losses have on the resolution of the multilayer superlens.

**2. The lens design**

For the present study we consider the superlens of Ref. 6 at a wavelength of 532 nm. The 1D MDPBG in Ref. 6 is composed of Ag/GaP and is super-resolving over the range of 500-650 nm while maintaining a transmittance for propagating waves of >50%. Fig. 1 shows the basic layout of the lens and the two source slits for the resolution and loss investigation. The multilayer stack contains five coupled Fabry-Perot cavities. The cavities are smaller than one-half wavelength due to the phase change upon reflection from thin metal films being considerably different from thick metal layers[10]. The lowest energy transmission resonance for a single Ag/GaP/Ag (22 nm/35 nm/22 nm) cavity in air occurs at a wavelength of 570 nm. In addition to the 5.5 periods of Ag/GaP there are GaP antireflection coatings on the end faces of the stack. Each GaP quarter-wave antireflection coating is 17 nm thick or one-half the thickness of the internal GaP layers[11]. The superlens is 341 nm thick and contains 39% volume of Ag. The distance between the source and image plane is 391 nm and the lens can be positioned from a=0 nm to a=50 nm (up against the source to a range of 50 nm away from the source).

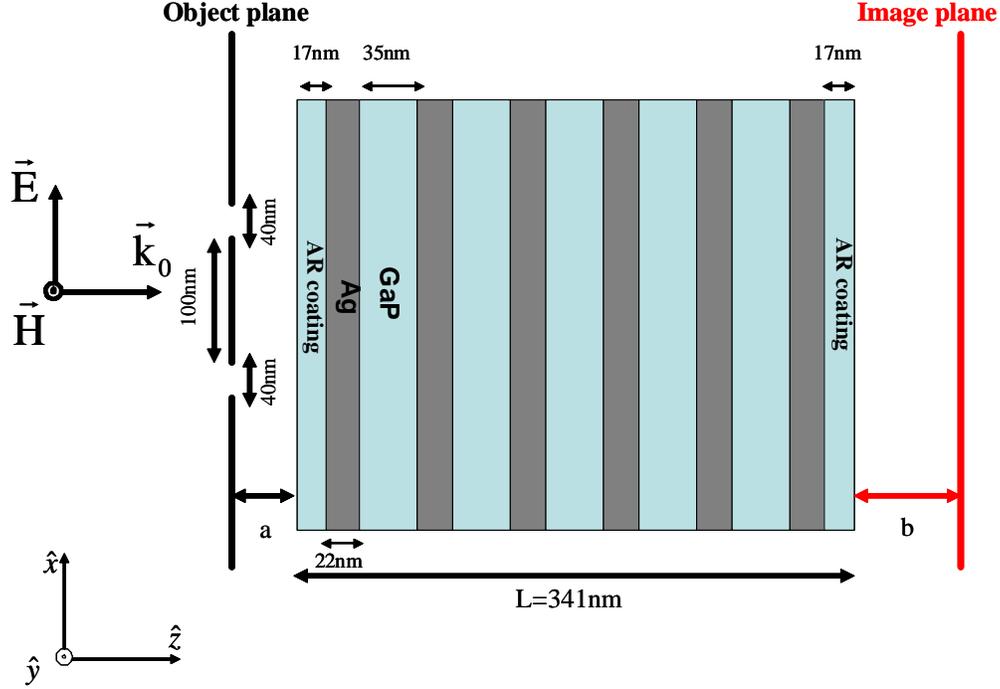

Fig. 1. Schematic representation of the lens, object plane with two slits, and the image plane. The lens consists of 5.5 periods of Ag/GaP (22 nm/35 nm) with GaP antireflection coatings, 17 nm thick, on the entrance and exit faces. The object plane to image plane distance is fixed at 391 nm and a+b=50 nm. A plane, monochromatic, TM polarized wave at a wavelength of $\lambda_0=532nm$ placed in vacuo is incident normally on the object plane, $k_0=2\pi/\lambda_0$ is the vacuum wave-vector. The strength of the electric and magnetic fields are in arbitrary units.

## 3. Theory

In this Section we detail on the calculation method used. We refer to TM polarization of the incident light. In our reference frame, the magnetic field diffracted is expressed as:

$$\vec{\tilde{H}}(x,z,t) = (1/2)\left[\vec{H}(x,z)\exp(-i\omega t) + c.c\right], \quad (1)$$

where $\vec{H}(x,z) = H(x,z)\hat{y}$ is the complex, stationary vector field, $\hat{y}$ is the unit vector of the y-axis, c.c. stands for complex conjugate, and the reference system (x,y,z) forms a right-handed Cartesian system. The complex amplitude of the magnetic field $H(x,z)$ is expressed through the angular spectrum decomposition technique [12]:

$$H(x,z) = \int_{-\infty}^{+\infty} A(k_x) H(k_x,z) \exp[ik_x x] dk_x, \quad (2)$$

where $k_x$ is a real quantity and it represents physically the wave-vector of the x-axis. $A(k_x)$ is the Fourier spectrum of the magnetic field on the object plane. In particular, in our case, $A(k_x)$ is the Fourier transform (FT) of the transmission function of the slits located at z=0:

$$A(k_x) = FT(t_{screen}(z=0,x)), \quad (3)$$

where the transmission function of the slits shown in Fig. 1 is defined as:

$$t_{screen}(z=0,x) = \begin{cases} 0 & -\infty < x < -D/2 - a_1 \\ 1 & -D/2 - a_1 \leq x \leq -D/2 \\ 0 & -D/2 < x < D/2 \\ 1 & D/2 \leq x \leq D/2 + a_2 \\ 0 & D/2 + a_2 < x < \infty \end{cases} \quad . \quad (4)$$

In practice, the transmission of the screen describes two slits respectively of width $a_1$ and $a_2$ located at a mutual distance (center to center) of $D+(a_1+a_2)/2$. Basically Eq.(4) is the sum of two rectangular functions. Eq.(4) applies to the idealized case of a perfectly absorbing, thin screen (black screen). In real experiments, the actual field values at the object plane will depend on the properties of the screen material and its thickness. An exact evaluation of these field's values would therefore require a detailed analysis of the interaction between the incident wave and the particular screen used. For example, transmission from nano-sized holes and slits in a metal screen is by itself a topic of great interest [13] whose description requires also some computational effort [14]. On the other hand, here we are interested in the description of the energy propagation through the multilayered lens more than how the incident wave interacts with the screen which is why we chose to use the simplest possible option, i.e. the black screen. (Based on numerical simulations of a realistic screen material with a finite thickness, we have found that the evanescent waves generated by a realistic screen decay more rapidly than those generated by the perfectly absorbing screen.) The Fourier transform of Eq.(4) can be performed analytically in terms of a linear superposition of sine cardinal ("sinc") functions. $H(k_x,z)$ are the Fourier modes over the x-axis that can be calculated by solving the following one-dimensional Helmholtz equation:

$$\frac{d^2 H(k_x,z)}{dz^2} + \left(\hat{n}(z)^2 k_0^2 - k_x^2\right) H(k_x,z) = 0, \quad (5)$$

where $k_0$ is the vacuum wave-vector, $\hat{n}(z)$ is the z-dependent, step-varying, complex refractive index of the different material layers present after the screen including also the possible air layer between the screen and the input surface of the lens and the air layer after the end surface of the lens. Eq.(5) can be solved analytically using a standard matrix transfer technique, Ref. 15. Once the integral in Eq.(2) has been calculated, the electric field can be calculated by using the differential form of the generalized Ampere law for time-harmonic fields in their complex representation:

$$\nabla \times \vec{H} = -i\omega\varepsilon(z)\vec{E}, \quad (6)$$

where $\varepsilon(z)$ is the z-dependent permittivity of the different material layers present after the screen including also the possible air layer between the screen and the input surface of the lens and the air layer after the end surface of the lens. The electric field obviously has two components both lying in the (x,z) plane $\vec{E}(x,z) = E_x(x,z)\hat{x} + E_z(x,z)\hat{z}$. Finally the Poynting vector is calculated by the well-known formula:

$$\vec{S} = (1/2)\text{Re}[\vec{E} \times \vec{H}^*]. \quad (7)$$

The rate of heat dissipated, i.e. the power dissipated, is calculated using the Landau Lifshitz formula [16], which in our case, given the fact that we are in a 2-D geometry and that we are considering non-magnetic materials, can be written as:

$$P^{dissipated} = \frac{\omega}{2} \iint Im(\varepsilon)\left[|E_x|^2 + |E_z|^2\right] dx dz \quad (8)$$

Although in this paper we have focused on TM polarization, the procedure outlined above can be easily applied to TE polarization with the formal substitution H→E into Eqs.(1,2,5) and using the differential form of Faraday's law of induction to calculate H from E. The same theoretical approach as the one explicitly described here has also already been used to study the formation of optical vortices during a super-resolution process [17].

**4. Transmittance of the lens**

Fig. 2 shows the transmittance of the lens for TM polarized light at a wavelength of 532 nm. The transmittance was calculated using the transfer matrix method as in Ref. 2. The optical constants for Ag ($n=0.13+i3.2$) and GaP ($n=3.5$) were taken from Palik[18]. We note that Ag films as thin as 20 nm have been fabricated in 1D MDPBGs with conventional deposition techniques [8-9]. The optical constants of bulk Ag were used to fit the transmission data in Refs 8-9 and agreed well with the measured transmittance spectra of the propagating waves. Quantum size effects in metals do not appear until the films are <3 nm thick (or about 6 lattice constants for Ag) since the de Broglie wavelength of electrons near the Fermi energy in metals is ~0.5 nm[19].

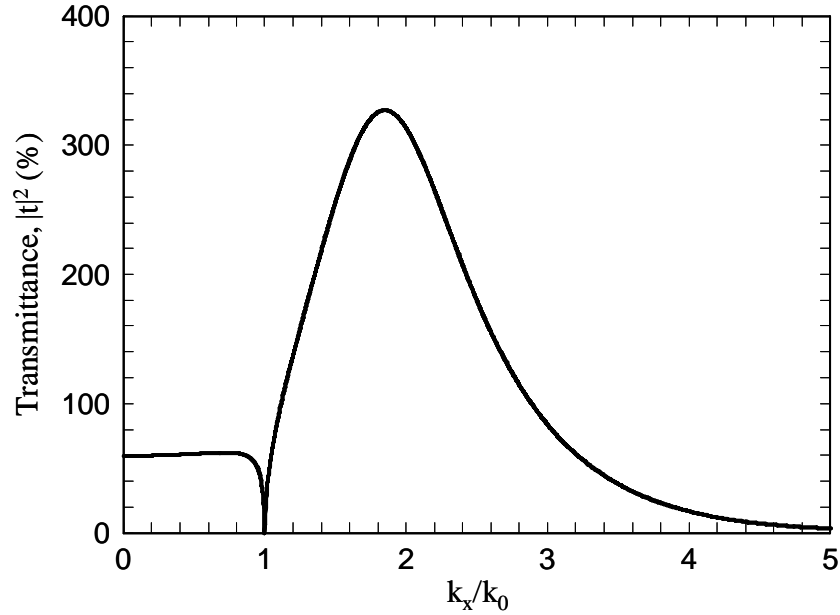

Fig. 2. Transmittance of the lens at a wavelength of 532 nm.

As shown in Fig. 2, the propagating waves have a transmittance of <100% while some of the evanescent waves have a transmittance >100%. Note that the transmittance for propagating waves is nearly constant except for large angles of incidence where $k_x/k_o$ approaches unity. This behavior is due to the large refractive index of GaP causing the rays to refract close to the normal for all angles of incidence. The result is a transparent metal having a passband without the usual blue-shift for increasing angles of incidence. A large refractive index for the dielectric in a MDPBG has other benefits related to the transmittance and superlensing properties. In general, a higher index dielectric reduces the reflectance at a metal/dielectric boundary due to a closer match in the magnitude of the real part of the dielectric constant. The quarter-wave antireflection coatings also substantially improve the transmittance for propagating waves [8]. The improved transmittance due to the high index dielectric is associated with a reduction in the reflectance and an increase in the absorbance. At 532 nm, the absorbance is 36% for the propagating waves.

The large refractive index of the dielectric has a significant impact on the evanescent waves as well [6]. The evanescent waves in metal/dielectric multilayers are associated with surface plasmons that propagate parallel to the metal/dielectric interfaces. Similar to the propagating waves, the high index dielectric allows for greater penetration into the metal resulting in increased absorption and a shift of the surface plasmon resonance towards larger values of the wavevector [20]. The broadening and shifting of the surface plasmon resonance is beneficial for the superlensing process. Superlenses with transmittance values of 10,000% for a narrow band of evanescent wavevectors are useful if the source spectrum is very narrow. The difficulty has been to design a lens with substantial transmittance over a broad range of Fourier components. Lossy surface plasmon modes can provide these broad transmission bands.

In the analysis of the absorption losses in the lens, we can separate the propagating and evanescent components. The two slits shown in Fig. 1 cause the propagating waves to diverge as they move away from the slit. Since the transmittance of the MDPBG depends weakly on the angle of incidence for the propagating waves we obtain results similar to the normal incidence case: the transmittance is 60%, the absorbance is 36% and the reflectance is 4%.

For the case of evanescent waves the transmittance has different connotations. While there is energy associated with evanescent waves, the energy does not propagate into the far field. The energy of the evanescent waves flows in the vicinity of the source and usually extends only several tens of nanometers beyond the slits. However, the spatial extent of the evanescent waves generated by the slits can be strongly influenced by a nearby resonant structure. A resonant structure which supports Fourier components found in the source spectrum can draw energy into it, affecting the spatial extent and energy density of the evanescent waves. If the resonant structure is lossy, it can draw energy from the source and dissipate the energy. The result is a net power flow into the resonator. The resonator can be an energy sink for the evanescent waves. This same basic idea has been proposed as a wireless method to recharge batteries for cameras and other portable devices [21].

The transmittance function for the evanescent waves relates the amplitude of the input wave to the output wave. For the evanescent waves the input amplitude is calculated without the lens in place. Therefore the input amplitude decreases as the lens is moved farther from the slits. On the output side of the resonant structure the evanescent waves resume their normal behavior and decay rapidly in free space. In the case of a lossless evanescent wave resonator, the transmittance can be >100% and yet no energy is dissipated. For a device with a transmittance >100% for propagating waves such as an amplifier, energy will be required to boost the wave amplitude.

In the superlensing process, the presence of the resonator structure extends the evanescent waves into what would normally be the far field, however, when the source and resonator are taken as a combined system, the near field has simply been extended to a greater distance.

## 5. Power flow in the lens

The effect of the 1D-MDPBG on the evanescent and propagating waves associated with two slits can be seen by plotting the z-component of the Poynting vector, $S_z$, in arbitrary units. Fig. 3 shows the free space case without the lens in place. At ~100 nm beyond the slits, the two slits have lost their identity. Note that the maximum value of $S_z$ is ~0.1 and this occurs very close to the slits. Note also that the total power emitted by the two slits and transported in the free space beyond the slits $P^{freespace}$, i.e. the z-component of the Poynting vector integrated over the transverse coordinate x, does not depend on the z-coordinate. The total power transported in free space $P^{free\,space} = \int_{-\infty}^{+\infty} S_z^{free\,space}(x,z)dx$ remains the same at any distance from the source, as one may expect. $P^{free\,space} \cong 8*10^{-3}$ is our normalization factor in arbitrary units.

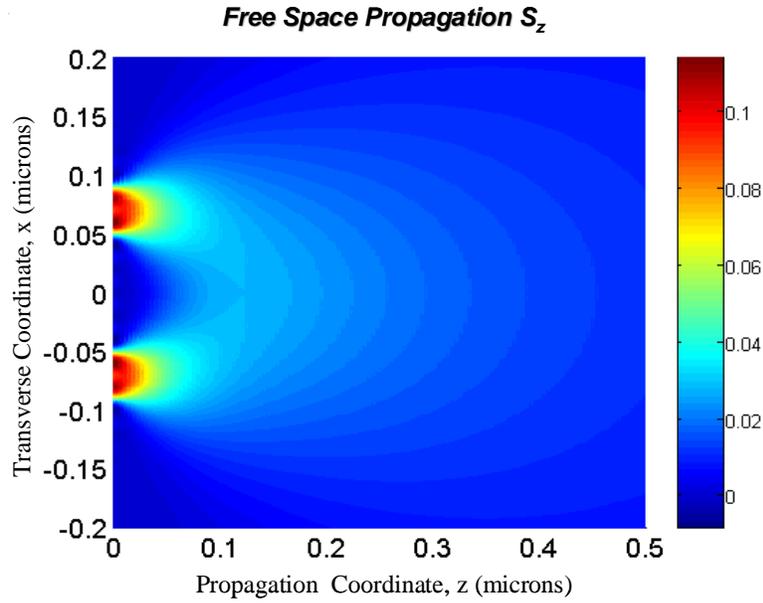

Fig. 3. Z-component of the Poynting vector (arbitrary units) in the space beyond the slits without the lens in place. The color is proportional to the power with red regions indicating maximum power flow. Note the maximum value of the color scale is 0.1.

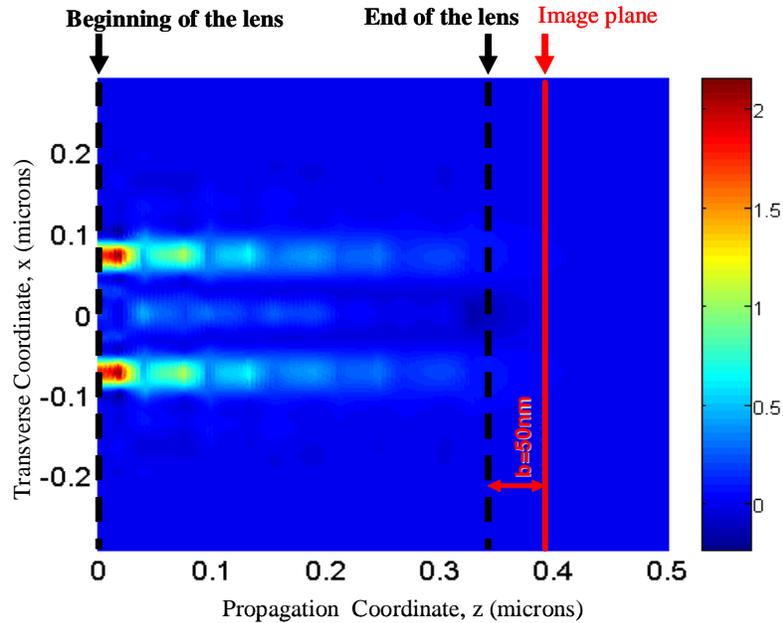

Fig. 4. Z-component of the Poynting vector in the space beyond the slits with the lens placed directly on the slits. The waveguiding by the lens is clearly seen by the two channels of optical power. Also notice that the maximum value of the color scale is 20 times higher than in Fig. 3 and indicates that the power flowing into the region downstream from the slits has increased dramatically due to the presence of the lens. The increased power flow results from the resonant coupling of the evanescent waves to the surface plasmon modes in the lens and the rapid dissipation of the surface plasmons through joule heating.

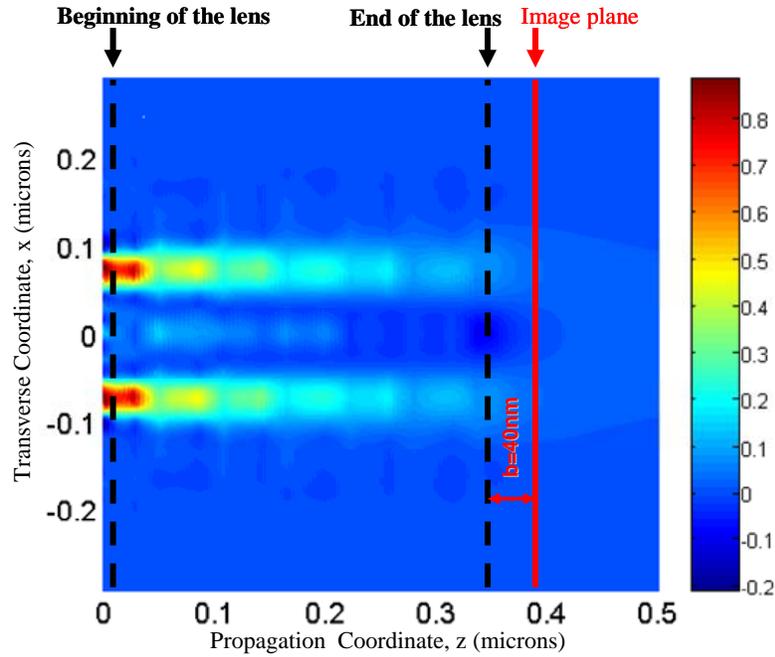

Fig. 5. Z-component of the Poynting vector propagating in the space beyond the slits with the lens placed 10 nm beyond the slits. The power flowing into the lens has been reduced by more than a factor of 2 in comparison to Fig. 4 due to the reduced coupling of the evanescent waves generated by the slits and the surface plasmon modes of the lens.

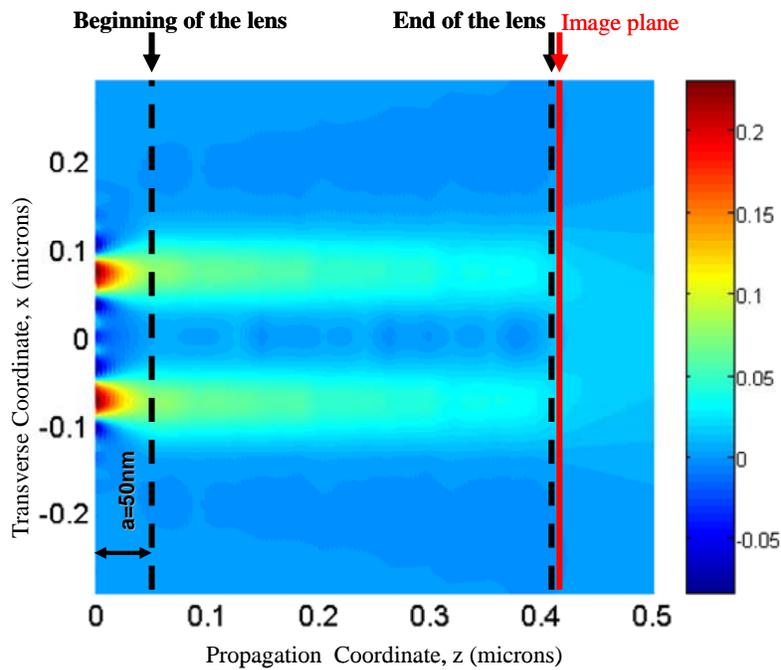

Fig. 6. Z-component of the Poynting vector propagating in the space beyond the slits with the lens placed 50 nm beyond the slits. The weaker coupling of the lens to the evanescent waves generated by the slit is evidenced by the maximum value of 0.2 for the color scale.

With the lens in place, Fig. 4, the two slits remain resolved at a distance of 400 nm away. In addition to the integrity of the image the other prominent feature in Fig. 4 is the magnitude of $S_z$ and the strong decay of $S_z$ as a function of distance from the slits. The scale in Fig. 4 is 20 times larger than compared to the case without the lens, Fig. 3.

The lens is drawing power from the source through the evanescent waves and the extra power is then dissipated in the Ag layers. Figs. 5 and 6 show $S_z$ for the lens positioned at 10 nm and 50 nm away from the slits respectively. For a 10 nm separation between the slits and the lens the maximum value of $S_z$ is 0.8 and for a 50 nm separation the maximum value of $S_z$ is only 0.2. Figs. 3-6 show that the power coupled into the lens decreases as the lens is placed progressively farther from the slits. Consequently, the power lost through the excitation of surface plasmons decreases as the distance between the lens and the slits expands

The losses associated with the placement of the lens can be visualized by plotting $S_z$ versus z where $S_z$ is integrated over the transverse coordinates ( $P(z) = \int_{-\infty}^{+\infty} S_z(x,z)dx$ ), Fig. 7. The graph illustrates the power lost through the successive Ag layers for various placements of the lens. The precipitous drop in the power flow along the direction of propagation is due to the absorption in the silver layers. The power flow inside each lossless layer of GaP is constant.

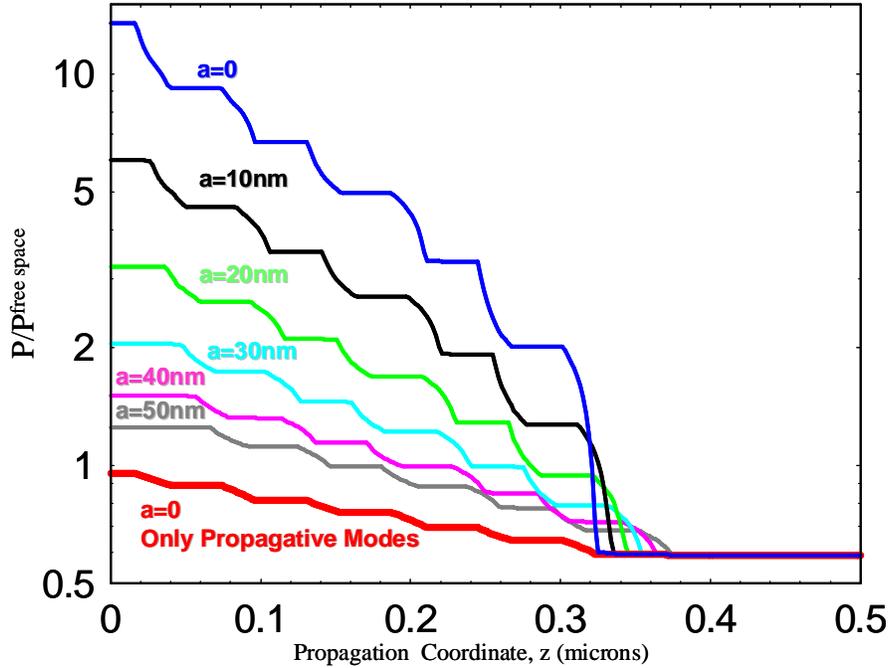

Fig. 7. Log-scale plot of the net power flowing through each layer of the lens for different placements of the lens with respect to the slits. The stair-step form of the plots indicates the power dissipated in the silver layers and not dissipated in the lossless GaP layers. The strength of the coupling of the evanescent waves to the lens is illustrated at the z=0 position. For the lens positioned at the slits, the maximum power is drawn into the lens and then dissipated in the successive Ag layers. Note that the power flowing beyond the lens is the same in every case. The net power flowing beyond the lens is attributed to the propagating waves only and is 60% of the initial power associated with the propagating waves. Beyond the lens, evanescent waves do not transfer power. Evanescent waves are present beyond the lens but do not carry a net power flow into the far field. Also shown is the case for only the propagating modes. The initial value of 96% and not 100% is due to the 4% reflectance for the propagating waves.

The strength of the coupling between the evanescent waves and the lens is indicated by looking at the power flowing just beyond the slit near z=0. For the case of a=0 (no separation between the lens and the slit) the coupling of the evanescent energy into the lens is a maximum. As the lens is moved away from the slits, the coupling strength for evanescent waves is reduced resulting in less power flowing into the lens.

Also shown in Fig. 7 is the case where the evanescent waves have been removed from the calculation. The case without evanescent components shows the usual behavior for propagating waves. Note at position z=0, the power flow is slightly less than unity due to the 4% reflectance of the MDPBG. The value of the power flowing beyond the end of the lens is 60% which is the transmittance of the lens.

Without the lens in place, no power is dissipated and the power flowing in the z-direction is unity for all values of z. The evanescent waves have energy associated with them but the net power flow along the z-axis without the lens in place is zero.

Fig. 8 shows the total power dissipated in the lens for different placements of the lens. The power dissipated has been normalized to what one would normally think of as the input power. This is the power incident on the lens that is associated with the propagating modes. When we only consider propagating waves the losses are independent of how far the lens is from the slits. When we add the losses associated with evanescent waves the power loss is seen to be strongly dependent on the placement of the lens. For the lens positioned at a=50 nm, the losses associated with the propagating waves are approximately the same value as the losses due to the evanescent waves. At a=0 nm the ohmic losses associated with the evanescent waves are ~30 times larger than losses associated with the propagating modes. This extra energy is associated with the excitation of low group velocity (stationary) surface plasmon waves which dissipate rapidly through ohmic heating.

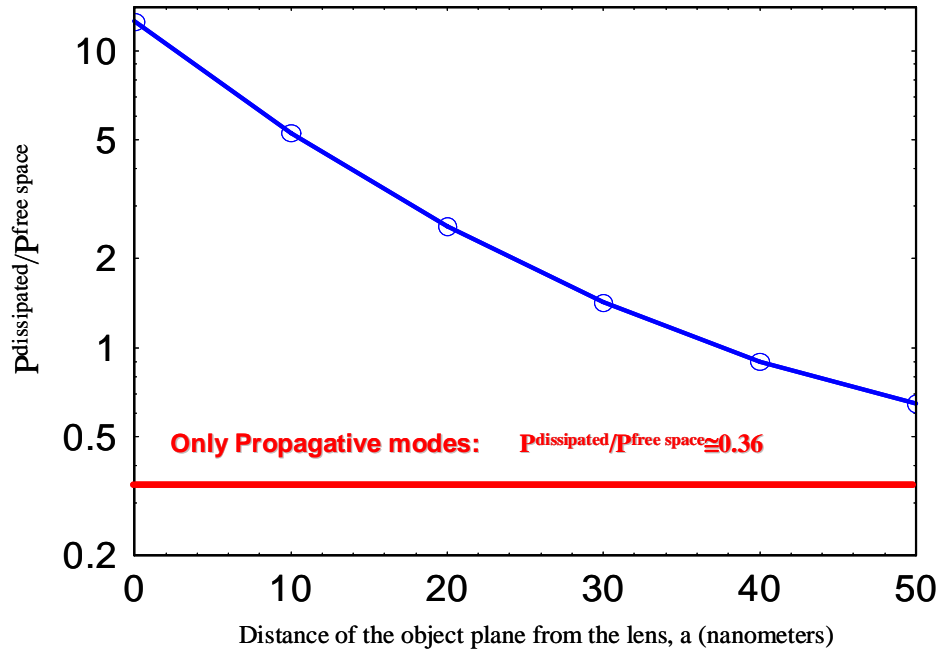

Fig. 8. Log-scale plot of the total power dissipated in the lens for different placements of the lens with respect to the slits. The red line in the plot is the case when the evanescent waves have been removed from the calculation. As expected the case for only the propagating waves shows a 36% power loss for every placement of the lens. This is consistent with the value of a 60% transmittance and 4% reflectance for the propagating waves. For the lens placed directly on the slits, the total power loss is dominated by the evanescent waves. For the lens placed 50 nm downstream from the slits the losses for the propagating waves and the evanescent waves are nearly equal.

Lastly we examine the resolution of the lens for an image plane situated at d=a+L+b=391 nm from the slits. Fig. 9 shows the two slits to be very well resolved at this distance. It was found that the exact placement of the lens between the slit plane and image plane has no effect on how well the two slits are resolved. The image resolution is independent of the lens placement. In Ref.[22] it had already been shown that for a generic negative index material, planar, super-lens, with Re($\varepsilon$)=Re($\mu$)=-1, the super-resolution was determined not by the distance between the image plane and the lens ,i.e. a, but rather by the total distance object plane to image plane, i.e. d=a+b+L. Normally, increased losses are considered detrimental to the resolution leading to the expectation of the resolution depending on the lens placement. In this case the losses can vary considerably without any change in the resolution. This result implies that the strength of the evanescent waves and the propagating waves at the position of the image do not depend on the position of the lens. The lens will amplify the evanescent components a fixed amount but the losses will be higher if the lens is placed near the slits where the evanescent fields are the strongest.

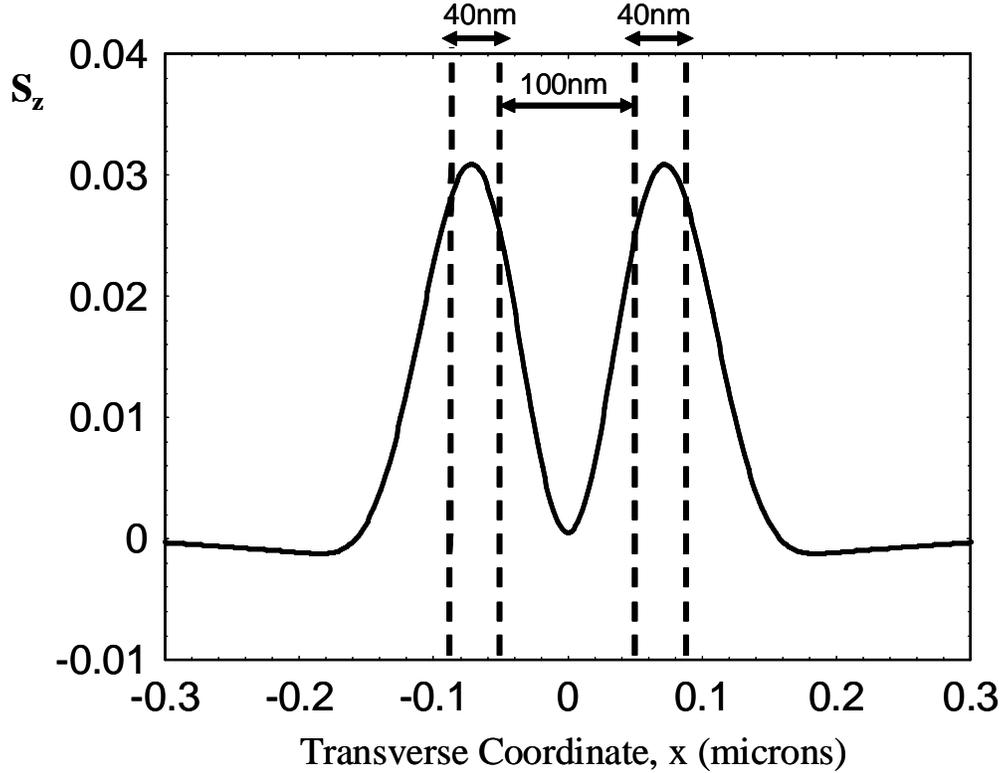

Fig. 9. Plotted is the z-component of the Poynting vector at the image plane. The total power flowing at the image plane and the resolution are independent of the position of the lens but depend on the total distance (d=L+a+b) image plane to object plane. Without the lens in place the slits are not resolved. The distance between the source and image plane was chosen somewhat arbitrarily to be the distance at which the z-component of the Poynting vector goes to zero at x=0. Due to vortices formed in the super resolution process, for d slightly less than 391 nm the z-component of the Poynting vector is negative and for d>391 nm the value is positive.

## 6. Conclusions

In conclusion we have analyzed the power flow in a realistic metallo-dielectric super-resolving lens. We have found that the losses associated with the evanescent waves in a superlensing process can far exceed the losses associated with the propagating waves. The losses associated with the evanescent modes strongly depend on the position of the lens with respect to the object plane while the losses associated with the propagative modes are independent of it. The net power transported at the output surface of the lens is the same with and without the evanescent modes and confirms the fact that the evanescent modes do not effectively carry energy into the far field. The presence of the lens causes power to flow through the evanescent fields into the lens and this "surplus" of energy is completely dissipated through joule heating in the metal layers so that in the far field the power remains the same. The overall loss of energy in the lens can be minimized by placing the lens close to the image plane without detrimental effects on the resolution. Although here we have analyzed explicitly a Ag/GaP multilayer, the same results are valid for metallo-dielectric super-resolving lenses based on different materials.